\documentclass[a4paper,authoryear,fleqn]{cas-dc}
\usepackage{natbib}

\usepackage{lipsum}

\xspaceaddexceptions{]}

\usepackage{multirow,booktabs,colortbl,tabularx}

\usepackage{amssymb}
\usepackage{xcolor}
\usepackage{hhline}

\usepackage[export]{adjustbox}% <-- added
\usepackage{caption}

\begin{document}

%%%%%%%%%%%%%%%%%%%%%%%%%%%%%%%%%%%%%%%%%%%%%%%
% Solve the issue with section and subsection numbering
%%%%%%%%%%%%%%%%%%%%%%%%%%%%%%%%%%%%%%%%%%%%%%%
\makeatletter
\renewcommand\section{\@startsection{section}{1}{\z@}%
    {15pt \@plus 3\p@ \@minus 3\p@}%
    {4\p@}%
    {%\let\@hangfrom\relax
     \sectionfont\raggedright\hst[13pt]}}
\renewcommand\subsection{\@startsection{subsection}{2}{\z@}%
    {10pt \@plus 3\p@ \@minus 2\p@}%
    {.1\p@}%
    {%\let\@hangfrom\relax
     \ssectionfont\raggedright }}
\renewcommand\subsubsection{\@startsection{subsubsection}{3}{\z@}%
    {10pt \@plus 1\p@ \@minus .3\p@}%
    {.1\p@}%
    {%\let\@hangfrom\relax
     \sssectionfont\raggedright}}
\makeatother
%%%%%%%%%%%%%%%%%%%%%%%%%%%%%%%%%%%%%%%%%%%%%%%

\let\WriteBookmarks\relax
\def\floatpagepagefraction{1}
\def\textpagefraction{.001}
\shorttitle{Deep-Learning-Based Audio-Visual Speech Enhancement in Presence of Lombard Effect}
\shortauthors{Michelsanti et al.}
%\begin{frontmatter}

\title [mode = title]{Deep-Learning-Based Audio-Visual Speech Enhancement in Presence of Lombard Effect}                      
%\tnotemark[1,2]

%%  \tnoteref{tn1,tn2}

%\tnotetext[1]{This document is the results of the research
%   project funded by the National Science Foundation.}

%\tnotetext[2]{The second title footnote which is a longer text matter
%   to fill through the whole text width and overflow into
%   another line in the footnotes area of the first page.}

\author[1]{Daniel Michelsanti}[orcid=0000-0002-3575-1600]
\cormark[1]
%\fnmark[1]
\ead{danmi@es.aau.dk}

\address[1]{Department of Electronic Systems, Aalborg University, Denmark}

\author[1]{Zheng-Hua Tan}[]
\ead{zt@es.aau.dk}

\author[2]{Sigurdur Sigurdsson}[]
%\fnmark[2]
\ead{ssig@oticon.com}

\address[2]{Oticon A/S, Denmark}

\author[1,2]{Jesper Jensen}
%\cormark[2]
%\fnmark[1,3]
\ead{jje@es.aau.dk, jesj@oticon.com}
%\ead[URL]{www.stmdocs.in}

\cortext[cor1]{Corresponding author}
%\fntext[fn1]{This is the first author footnote. but is common to third
%  author as well.}
%\fntext[fn2]{Another author footnote, this is a very long footnote and
%  it should be a really long footnote. But this footnote is not yet
%  sufficiently long enough to make two lines of footnote text.}

%\nonumnote{This note has no numbers. In this work we demonstrate $a_b$
%  the formation Y\_1 of a new type of polariton on the interface
%  between a cuprous oxide slab and a polystyrene micro-sphere placed
%  on the slab. The evanescent field of the resonant whispering gallery
%  mode (\WGM) of the micro sphere has a substantial gradient, and
%  therefore effectively couples with the quadrupole $1S$ excitons in
%  cuprous oxide.
%  \lipsum[1]
%  }

\begin{abstract}[S U M M A R Y]

When speaking in presence of background noise, humans reflexively change their way of speaking in order to improve the intelligibility of their speech. This reflex is known as Lombard effect. Collecting speech in Lombard conditions is usually hard and costly. For this reason, speech enhancement systems are generally trained and evaluated on speech recorded in quiet to which noise is artificially added. Since these systems are often used in situations where Lombard speech occurs, in this work we perform an analysis of the impact that Lombard effect has on audio, visual and audio-visual speech enhancement, focusing on deep-learning-based systems, since they represent the current state of the art in the field.

We conduct several experiments using an audio-visual Lombard speech corpus consisting of utterances spoken by $54$ different talkers. The results show that training deep-learning-based models with Lombard speech is beneficial in terms of both estimated speech quality and estimated speech intelligibility at low signal to noise ratios, where the visual modality can play an important role in acoustically challenging situations. We also find that a performance difference between genders exists due to the distinct Lombard speech exhibited by males and females, and we analyse it in relation with acoustic and visual features. Furthermore, listening tests conducted with audio-visual stimuli show that the speech quality of the signals processed with systems trained using Lombard speech is statistically significantly better than the one obtained using systems trained with non-Lombard speech at a signal to noise ratio of $-5$ dB. Regarding speech intelligibility, we find a general tendency of the benefit in training the systems with Lombard speech.
\end{abstract}

\maketitle

\section{Introduction}

\textit{Speech} is perhaps the most common way that people use to communicate with each other. Often, this kind of communication is harmed by several sources of disturbance that may have different nature, such as the presence of competing speakers, the loud music during a party, and the noise inside a car cabin. We refer to the sounds other than the speech of interest as \textit{background noise}.

Background noise is known to affect two attributes of speech: \textit{intelligibility} and \textit{quality} \citep{loizou2007speech}. Both of these aspects are important in a conversation, since poor intelligibility makes it hard to comprehend what a speaker is saying and poor quality may affect speech naturalness and listening effort \citep{loizou2007speech}. Humans tend to tackle the negative effects of background noise by instinctively changing the way of speaking, their \textit{speaking style}, in a process known as \textit{Lombard effect} \citep{lombard1911signe, zollinger2011evolution}. The changes that can be observed vary wide\-ly across individuals \citep{junqua1993lombard, marxer2018impact} and  affect multiple dimensions: acoustically, the average fundamental frequency (F0) and the sound energy increase, the spectral tilt flattens due to an energy increment at high frequencies and the centre frequency of the first and second formant (F1 and F2) shifts \citep{junqua1993lombard, lu2008speech}; visually, head and face motion are more pronounced and the movements of the lips and jaw are amplified \citep{vatikiotis2007audiovisual, garnier2010influence, garnier2012effect}; temporally, the speech rate changes due to an increase of the vowel duration \citep{junqua1993lombard, cooke2014listening}.

Although Lombard effect improves the intelligibility of speech in noise \citep{summers1988effects, pittman2001recognition}, effective communication might still be challenged by some particular conditions, e.g. the hearing impairment of the listener. In these situations, \textit{speech enhancement} (SE) algorithms may be applied to the noisy signal aiming at improving speech quality and speech intelligibility. In the literature, several SE techniques have been proposed. Some approaches consider SE as a \textit{statistical estimation} problem \citep{loizou2007speech}, and include some well-known methods, like the Wiener filtering \citep{lim1979enhancement} and the minimum mean square error estimator of the short-time magnitude spectrum \citep{ephraim1984speech}. Many improv\-ed methods have been proposed, which primarily distinguish themselves by refined statistical speech models \citep{martin2005speech, erkelens2007minimum, gerkmann2009statistics} or noise models \citep{martin2003speech, loizou2007speech}. These techniques, which make statistical assumptions on the distributions of the signals, have been reported to be largely unable to provide speech intelligibility improvements \citep{hu2007comparative, jensen2012spectral}. As an alternative, \textit{data-driven techniques}, especially deep learning, do not make any assumptions on the distribution of the speech, of the noise or on the way they are mixed: a learning algorithm is used to find a function that best maps features from degraded speech to features from clean speech. Over the years, the speech processing community has put a considerable effort into designing training targets and objective functions \citep{wang2014training, erdogan2015phase, williamson2016complex, michelsanti2018training} for different neural network models, including deep neural networks \citep{xu2014experimental, kolbk2017speech}, denoising autoencoders \citep{lu2013speech}, recurrent neural networks \citep{weninger2014discriminatively}, fully convolutional neural networks \citep{park2016fully}, and generative adversarial networks \citep{michelsanti2017conditional}. These methods represent the current state of the art in the field \citep{wang2018supervised}, and since they use only audio signals, we refer to them as audio-only SE (AO-SE) systems.

Previous studies show that observing the speaker's facial and lip movements contributes to speech perception \citep{sumby1954visual, erber1975auditory, mcgurk1976hearing}. This finding suggests that a SE system could tolerate higher levels of background noise, if visual cues could be used in the enhancement process. This intuition is confirmed by a pioneering study on audio-visual SE (AV-SE) by \citet{girin2001audio}, where simple geometric features extracted from the video of the speaker's mouth are used. Later, more complex frameworks based on classical statistical approaches ha\-ve been proposed \citep{almajai2011visually, abel2014novel, abel2014cognitively}, and very recently deep learning methods have been used for AV-SE \citep{hou2017audio, gabbay2017visual, ephrat2018looking, afouras2018conversation, owens2018audio, morrone2018face}.

It is reasonable to think that visual features are mostly helpful for SE when the speech is so degraded that AO-SE systems achieve poor performance, i.e. when background noise heavily dominates over the speech of interest. Since in such acoustical environment spoken communication is particularly hard, we can assume that the speakers are under the influence of Lombard effect. In other words, the input to SE systems in this situation is Lombard speech. Despite this consideration, state-of-the-art SE systems do not take Lombard effect into account, because collecting Lombard speech is usually expensive. The training and the evaluation of the systems are usually performed with speech recorded in quiet and afterwards degraded with additive noise. Previous work shows that speaker \citep{hansen2009analysis} and speech recognition \citep{junqua1993lombard} systems that ignore Lombard effect achieve sub-optimal performance, also in visual \citep{heracleous2013analysis, marxer2018impact} and audio-visual settings \citep{heracleous2013analysis}. It is therefore of interest to conduct a similar study also in a SE context.

With the objective of providing a more extensive analysis of the impact of Lombard effect on deep-learning-based SE systems, the present work extends a preliminary study \citep{michelsanti2018effects}, providing the following novel contributions. First, new experiments are conducted, where deep-learning-based SE systems trained with Lombard or non-Lombard speech are evaluated on Lombard speech using a cross-validation setting to avoid that a potential intra-speaker variability of the adopted dataset leads to biased conclusions. Then, an investigation of the effect that the inter-speaker variability has on the systems is carried out, both in relation to acoustic as well as visual features. Next, as an example application, a system trained with both Lombard and non-Lombard data using a wide signal-to-noise-ratio (SNR) range is compared with a system trained only on non-Lombard speech, as it is currently done for the state-of-the-art models. Finally, especially since existing objective measures are limited to predict speech quality and intelligibility from the audio signals in isolation, listening tests using audio-visual stimuli have been performed. This test setup, which is generally not employed to evaluate SE systems, is closer to a real-world scenario, where a listener is usually able to look at the face of the talker.

\section{Materials: Audio-Visual Speech Corpus and Noise Data}
\label{sec:materials}

The speech material used in this study is the Lombard GRID corpus \citep{alghamdi2018corpus}, which is an extension of the popular audio-visual GRID dataset \citep{cooke2006audio}. It consists of 55 native speakers of British English ($25$ males and $30$ females) that are between $18$ and $30$ years old. The sentences pronounced by the talkers adhere to the syntax from the GRID corpus, six-word sentences with the following structure: <command> <color*> <preposition> <letter*> <digit*> <adverb> (Table \ref{tab:syntax}). The words marked with a * are keywords, whereas the others are fillers \citep{cooke2006audio}.

\begin{table}
\centering
\resizebox{0.47\textwidth}{!}{%
\begin{tabular}{c c c c c c }
\toprule
Command & Colour* & Preposition & Letter* & Digit* & Adverb \\
\midrule
bin & blue & at & & \multirow{ 4}{*}{0--9} & again \\
lay & green & by & A--Z & & now \\
place & red & in & (no W) & & please \\
set & white & with & & & soon \\
\bottomrule
\end{tabular}}
\caption{Sentence structure for the Lombard GRID corpus \citep{alghamdi2018corpus}. The `*' indicates a keyword. Adapted from \citep{cooke2006audio}.}
\label{tab:syntax}
\end{table}

Each speaker was recorded while reading a unique set of $50$ sentences in non-Lombard (NL) and Lombard (L) conditions (in total, $100$ utterances per speaker). In both cases, the audio signals were recorded with a microphone placed in front of the speakers, while the video recordings were collected with two cameras mounted on a helmet to have a frontal and a profile views of the talkers.

In order to induce the Lombard effect, speech shaped noise (SSN) at $80$ dB sound pressure level (SPL) was presented to the speakers, while they were reading the sentences to a listener. The presence of a listener, who assured a natural communication environment by asking the participants to repeat the utterances from time to time, was needed, because talkers usually adjust their speech to communicate better with the people they are talking to \citep{lane1971lombard, lu2008speech}, a process known as \textit{external} or \textit{public loop} \citep{lane1971lombard}. Since talkers tend to regulate their speaking style also based on the level of their own speech, in what is generally called \textit{internal} or \textit{private loop} \citep{lane1971lombard}, the speech signal was mixed with the SSN at a carefully adjusted level, providing a self-monitoring feedback to the speakers. 

In our study, the audio and the video signals from the frontal camera were arranged as explained in Section \ref{sec:experiments} to build training, validation, and test sets. The audio signals have a sampling rate of $16$ kHz. The resolution of the frontal video stream is $720\times480$ pixels with a variable frame rate of around $24$ frames per second (FPS). Audio and video signals are temporally aligned.

To generate speech in noise, SSN was added to the audio signals of the Lombard GRID database. SSN was chosen to match the kind of noise used in the database, since, as reported by \citet{hansen2009analysis}, Lombard effect occurs differently across noise types, although other studies \citep{lu2009speech, garnier2014speaking} failed to find such an evidence. The SSN we used was generated as in \citep{kolboek2016speech}, by filtering white noise with a low-order linear predictor, whose coefficients were found using 100 random sentences from the Akustiske Databaser for Dansk (ADFD)\footnote{\url{https://www.nb.no/sbfil/dok/nst_taledat_dk.pdf}} speech database.

\section{Methodology}

\begin{figure*}
	\centering
		\includegraphics[scale=.19]{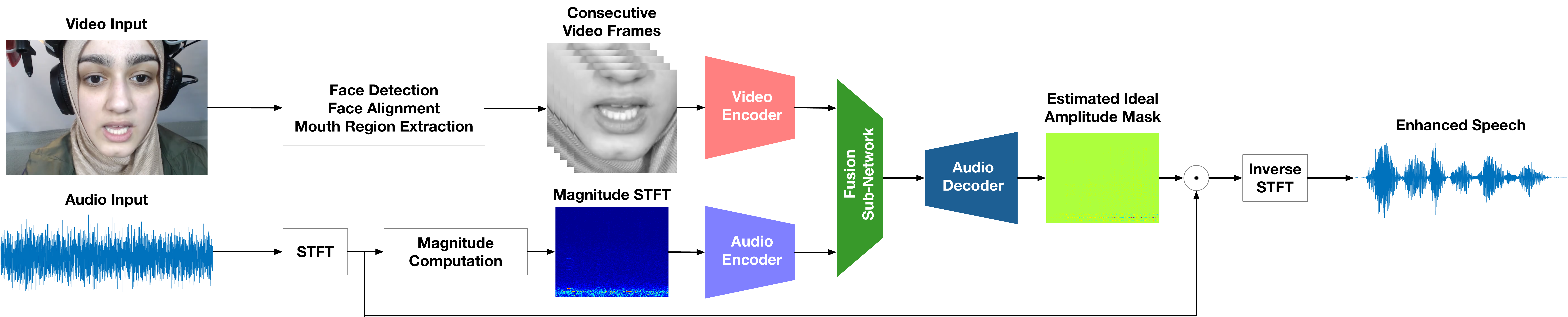}
	\caption{Pipeline of the audio-visual speech enhancement framework used in this study, adapted from \citep{gabbay2017visual}, and identical to \citep{michelsanti2018effects}. The deep-learning-based system estimates an ideal amplitude mask from the video of the speaker's mouth and the magnitude spectrogram of the noisy speech. The estimated mask is used to enhance the speech in time-frequency domain. \textit{STFT} indicates the short-time Fourier transform.}
	\label{fig:framework}
\end{figure*}

In this study, we train and evaluate systems that perform spectral SE using deep learning, as illustrated in Figure~\ref{fig:framework}. The processing pipeline is inspired by \citet{gabbay2017visual} and the same as the one used in \citep{michelsanti2018effects}. To have a self-contained exposition, we report the main details of it in this section.

\subsection{Audio-Visual Speech Enhancement}

We assume to have access to two streams of information: the video of the talker's face, and an audio signal, $y(n) = x(n) + d(n)$, where $x(n)$ is the clean signal of interest, $d(n)$ is an additive noise signal, and $n$ indicates  the discrete-time index. The additive noise model presented in time domain, can also be expressed in the time-frequency (TF) domain as $Y(k, l) = X(k, l) + D(k, l)$, where $Y(k, l)$, $X(k, l)$, and $D(k, l)$ are the short-time Fourier transform (STFT) coefficients at frequency bin $k$ and at time frame $l$ of $y(n)$, $x(n)$, and $d(n)$, respectively. Our models adopt a mask approximation approach \citep{michelsanti2018training}, producing an estimate $\widehat{M}(k,l)$ of the ideal amplitude mask, defined as $M(k,l) = | X(k,l) |/| Y(k,l) |$, with the following objective function:
\begin{equation}J = \frac{1}{TF} \sum_{k,l} \big( M(k,l) - \widehat{M}(k,l)\big)^2, \label{eqn:IAM-MA}\end{equation}
\noindent with $k \in \{1, \ldots , F\}$, $l \in \{ 1, \ldots , T\}$, and $T \times F$ being the dimension of the training target. Recent preliminary experiments have shown that using this objective function leads to better performance for AV-SE than competing methods \citep{michelsanti2018training}.

\subsection{Preprocessing}

In this work, each audio signal was peak-normalised. We used a sample rate of $16$ kHz and a $640$-point STFT, with a Hamming window of $640$ samples ($40$ ms) and a hop size of $160$ samples ($10$ ms). Only the 321 bins that cover the positive frequencies were used, because of the conjugate symmetry of the STFT.

Each video signal was resampled at a frame rate of $25$ FPS using motion interpolation as implemented in FFMPEG\footnote{\url{http://ffmpeg.org}}. The face of the speaker was detected in every frame using the frontal face detector implemented in the dlib toolkit \citep{dlib09}, consisting of $5$ histogram of oriented gradients (HOG) filters and a linear support vector machine (SVM). The bounding box of the single-frame detections was tracked using a Kalman filter. The face was aligned based on $5$ landmarks using a model that estimated the position of the corners of the eyes and of the bottom of the nose \citep{dlib09} and was scaled to $256\times256$ pixels. The mouth was extracted by cropping the central lower face region of size $128\times128$ pixels.

Each segment of $5$ consecutive grayscale video frames spanning a total of $200$ ms was paired with the respective $20$ consecutive audio frames. 

\subsection{Neural Network Architecture and Training}

The preprocessed audio and video signals, standardised using the mean and the variance from the training set, were used as input to a video and an audio encoders, respectively. Both encoders consisted of $6$ convolutional layers, each of them followed by leaky-ReLU activation functions \citep{maas2013rectifier} and batch normalisation \citep{ioffe2015batch}. For the video encoder, also max-pooling and 0.25 dropout \citep{hinton2012improving} were adopted. The fusion of the two modalities was accomplished using a sub-network consisting of 3 fully connected layers, followed by leaky-ReLU activations, on the outputs of the 2 encoders. The $321\times20$ estimated mask was obtained with an audio decoder having 6 transposed convolutional layers followed by leaky-ReLU activations and a ReLU activation as output layer. Skip connections between the layers $1$, $3$, and $5$ of the audio encoder and the corresponding decoder layers were used to avoid that the bottleneck hindered the information flow \citep{isola2017image}. The values of the training target, $M(k,l)$, were limited in the $[0, 10]$ interval \citep{wang2014training}. 

The weights of the network were initialised with the Xa\-vier approach \citep{glorot2010understanding}. The training was performed using the Adam optimiser \citep{kingma2014adam} with the objective function in Equation (\ref{eqn:IAM-MA}) and a batch size of $64$. The learning rate, initially set to $4\cdot10^{-4}$, was scaled by a factor of $0.5$ when the loss increased on the validation set. An early stopping technique was used, by selecting the network that performed the best on the validation set across the $50$ epochs used for training.

\subsection{Postprocessing}

The estimated ideal amplitude mask of an utterance was obtained by concatenating the outputs of the network, obtained by processing non-overlapping consecutive audio-vi\-sual paired segments. The estimated mask was point-wise multiplied with the complex-valued STFT spectrogram of the noisy signal and the result inverted using an overlap-add procedure to get the time-domain signal \citep{allen1977short, griffin1984signal}.

\subsection{Mono-Modal Speech Enhancement}

Until now, we only presented AV-SE systems. In order to understand the relative contribution of the audio and the visual modalities, we also trained networks to perform mono-modal SE, by removing one of the two encoders from the neural network architecture, without changing the other explained settings and procedures. Both AO-SE and video-only SE (VO-SE) systems estimate a mask and apply it to the noisy speech, but they differ in the signals used as input.

\section{Experiments}
\label{sec:experiments}

The experiments conducted in this study compare the performance of AO-SE, VO-SE, and AV-SE systems in terms of two widely adopted objective measures: perceptual evaluation of speech quality (PESQ) \citep{rix2001perceptual}, specifically the wideband extension \citep{itu2005recommendation} as implemented by \citet{loizou2007speech}, and extended short-time objective intelligibility (ESTOI) \citep{jensen2016algorithm}. PESQ scores, used to estimate speech quality, lie between $-0.5$ and $4.5$, where high values correspond to high speech quality. However, the wideband extension that we use maps these scores to mean opinion score (MOS) values, on a scale from approximately 1 to 4.64. ESTOI scores, which estimate speech intelligibility, practically range from 0 to 1, where high values correspond to high speech intelligibility.

As mentioned before (Section \ref{sec:materials}), clean speech signals were mixed with SSN to match the noise type used in the Lombard GRID corpus. Current state-of-the-art SE systems are trained with signals at several SNRs to make them robust to various noise levels. We followed a similar methodology and trained our models with two different SNR ranges, narrow (between $-20$ dB and $5$ dB) and wide (between $-20$ dB and $30$ dB). We used these two ranges because on the one hand we would like to assess the performance of SE systems when Lombard speech occurs, and on the other hand we would like to have SNR-independent systems, i.e. systems that also work well at higher SNRs. Such a setup allows us to better understand whether Lombard speech, which is usually not available because it is hard to collect, should be used to train SE systems and which are the advantages and the disadvantages of various training configurations. The models used in this work are shown in Table \ref{tab:models}.

\begin{table}
\centering
\resizebox{0.48\textwidth}{!}{
\begin{tabular}{l | c c | c c}
\toprule
&\multicolumn{4}{c}{\begin{tabular}{@{}c@{}}Training Material\end{tabular}}\\
\midrule
 & \multicolumn{2}{c}{\begin{tabular}{@{}c@{}}Non-Lombard \\ Speech\end{tabular}} & \multicolumn{2}{c}{\begin{tabular}{@{}c@{}}Lombard \\ Speech\end{tabular}} \\
 \begin{tabular}{@{}l@{}}System \\ Input \end{tabular}  & \begin{tabular}{@{}c@{}}Narrow \\ SNR Range \end{tabular} & \begin{tabular}{@{}c@{}}Wide \\ SNR Range \end{tabular} & \begin{tabular}{@{}c@{}}Narrow \\ SNR Range \end{tabular} & \begin{tabular}{@{}c@{}}Wide \\ SNR Range \end{tabular} \\
\midrule
Vision     		&    VO-NL & VO-NL\textsuperscript{(w)} &    VO-L & VO-L\textsuperscript{(w)}        \\ 
Audio      		&    AO-NL & AO-NL\textsuperscript{(w)} &    AO-L & AO-L\textsuperscript{(w)}       \\ 
Audio-Visual     	&    AV-NL & AV-NL\textsuperscript{(w)}   &    AV-L & AV-L\textsuperscript{(w)}    \\ 
\bottomrule
\end{tabular}}
\caption{Models used in this study. The `\textsuperscript{(w)}'  is used to distinguish the systems trained with a wide SNR range from the ones trained with a narrow SNR range.}
\label{tab:models}
\end{table}

Similarly to the work by \citet{marxer2018impact}, the experiments were conducted adopting a multi-speaker setup, in which all the speakers in the database were used for both training and evaluating the systems. This choice was made for a practical reason. People may exhibit speech characteristics that differ considerably from each other when  they speak in presence of noise \citep{junqua1993lombard, marxer2018impact}. It is possible to model these differences by training speaker-dependent systems, but this requires a large set of Lombard speech for every speaker. Unfortunately, the audio-visual speech corpus that we use, despite being one of the largest existing audio-visual databases for Lombard speech, only contains $50$ utterances per speaker, which are not enough to train a deep-learning-based model.

The experiments were performed according to a stratified five-fold cross-validation procedure \citep{liu2009encyclopedia}. Specifically, the data was divided into five folds of approximately the same size, four of them used for training and validation, and one for testing. This process was repeated five times for different test sets in order to evaluate the systems on the whole dataset. Before the split, the signals were rearranged to have about the same amount of data for each speaker across the training ($\sim35$ utterances), the validation ($\sim5$ utterances), and the test ($\sim10$ utterances) sets. This ensured that each fold was a good representative of the inter-speaker variations of the whole dataset. For some speakers, some data was missing or corrupted, so we used fewer utterances. Among the $55$ speakers, the recordings from speaker s$1$ were discarded by the database collectors due to technical issues, and the data from speaker s$51$ was used only in the training set, because only 40 of the utterances could be used. Effectively, $53$ speakers were used to evaluate our systems.

\subsection{Systems Trained on a Narrow SNR Range}
\label{sec:narrowsnrmodels}

Since we would like to assess the performance of SE systems when Lombard speech occurs, SSN is added to the speech signals from the Lombard GRID corpus at 6 different SNRs, in uniform steps between $-20$ dB and $5$ dB. This choice was driven by the following considerations \citep{michelsanti2018effects}. Since Lombard and non-Lombard utterances from the Lombard GRID corpus have an energy difference between $3$ and $13$ dB \citep{marxer2018impact}, the actual SNR can be computed assuming that the conversational speech level is between $60$ and $70$ dB sound pressure level (SPL) \citep{raphael2007speech, moore2012introduction} and the noise level at $80$ dB SPL, like in the recording conditions of the database. The SNR range obtained in this way is between $-17$ and $3$ dB. In the experiments, we used a slightly wider range because of the possible speech level variations caused by the distance between the listener and the speaker. 

For all the systems, Lombard speech was used to build the test set, while for training and validation we used Lombard speech for VO-L, AO-L, and AV-L, and non-Lombard speech for VO-NL, AO-NL, and AV-NL (Table \ref{tab:models}).

\subsubsection{Results and Discussion}
\label{sec:narrowres}

\begin{figure*}
	\centering 
		\includegraphics[scale=.52]{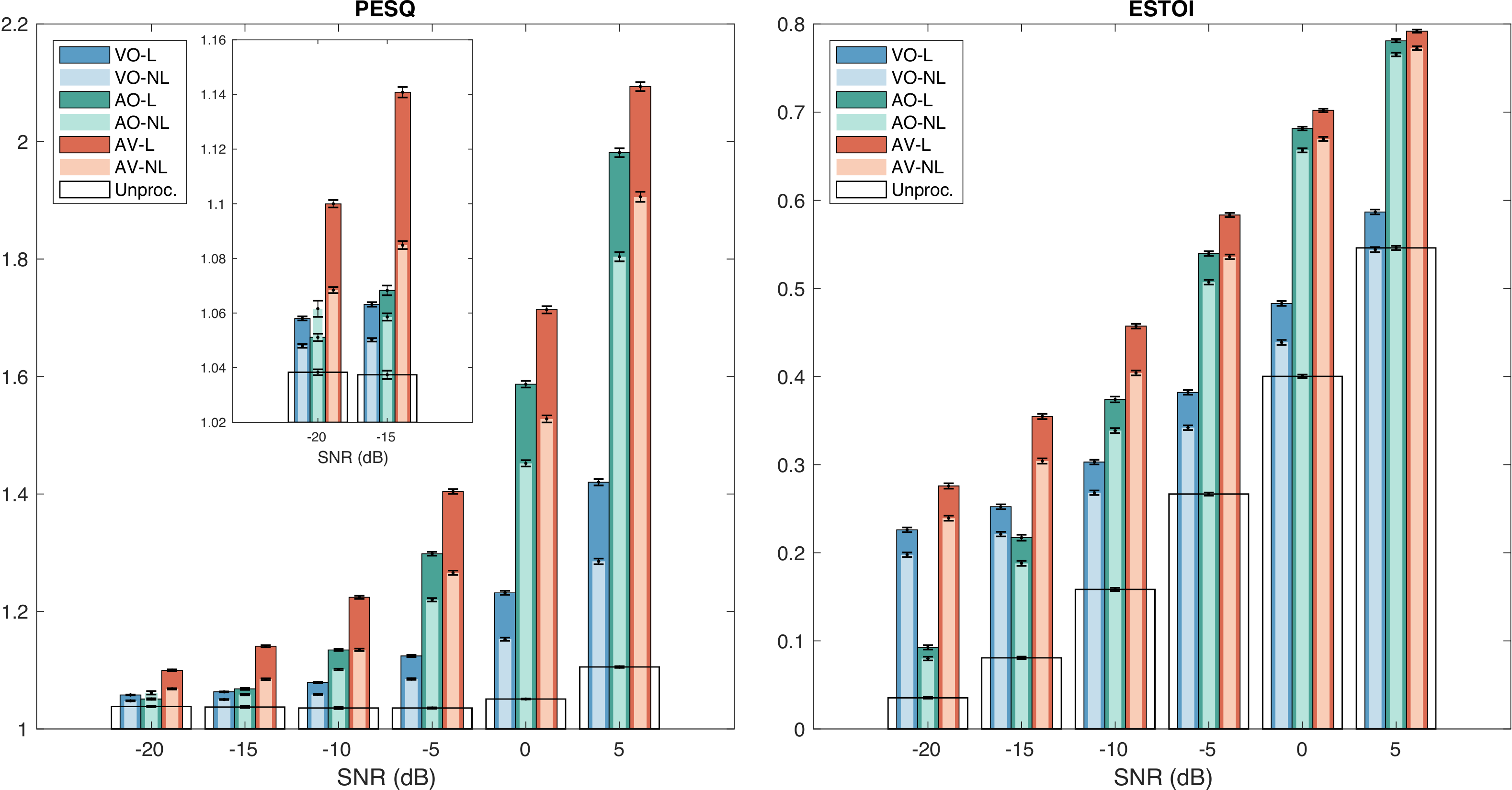}
	\caption{Cross-validation results in terms of PESQ and ESTOI for the systems trained on a narrow SNR range. At every SNR, there are three pairs of coloured bars with error bars, each of them referring to VO, AO, and AV systems (from left to right). The wide bars in dark colours represent L systems, while the narrow ones in light colours represent NL systems. The heights of each bar and the error bars indicate the average scores and the 95\% confidence intervals computed on the pooled data, respectively. The transparent boxes with black edges, overlaying the bars of the other systems, and the error bars indicate the average scores of the unprocessed signals (\textit{Unproc.}) and their 95\% confidence intervals, respectively.}
	\label{fig:res_seen}
\end{figure*}

Figure \ref{fig:res_seen} shows the cross-validation results in terms of PESQ and ESTOI for all the different systems. On average, every model improves the estimated speech quality and the estimated speech intelligibility of the unprocessed signals, with the exception of VO-NL at 5 dB SNR, which shows an ESTOI score comparable with the one of noisy speech. Another general trend that can be observed is that AV systems outperform the respective AO and VO systems, an expected result since the information that can be exploited using two modalities is no less than the information of the single modalities taken separately.

It is worth noting that VO systems' performance changes across SNR, although they do not use the audio signal to estimate the ideal amplitude mask. This is because the estimated mask is applied to the noisy input signal, so the performance depends on the noise level of the input audio signal.

PESQ scores show that the performance that can be obtained with AO systems is comparable with VO systems performance at very low SNRs. Only for $\text{SNR} \geq -10$ dB, AO models start to perform substantially better than VO models. The difference increases with higher SNRs. Also for ESTOI, this pattern can be observed when $\text{SNR} \geq -10$ dB, but for $\text{SNR} \leq -15$ dB VO systems perform better than the respective AO systems, especially at $-20$ dB SNR where the performance gap is very large. This can be explained by the fact that the noise level is so high that recovering the clean speech only using the noisy audio input is very challenging, and that the visual modality provides a richer information source at this noise level.

For all the modalities, L systems tend to be better than the respective NL systems. The only exception is AO-NL, which have a higher PESQ score than AO-L at $-20$ dB SNR, but this difference is very modest ($0.011$).  AV-L always outperforms AV-NL in terms of PESQ by a large margin, with more than $5$ dB SNR gain, if we consider the performance between $-20$ dB and $-10$ dB SNR. On average (Table \ref{tab:res_nar}), the performance gap in terms of PESQ between L and NL systems, is greater for the audio-visual case ($0.115$) than for the audio-only ($0.070$) and the video-only ($0.050$) cases, meaning that the speaking style mismatch is more detrimental when both the modalities are used. Regarding ESTOI, the gap between AV-L and AV-NL ($0.040$) is still the largest, but the one between VO-L and VO-NL ($0.037$) is greater than the gap between AO-L and AO-NL ($0.025$): this suggests that the impact of visual differences between Lombard and non-Lombard speech on estimated speech intelligibility is higher than the impact of acoustic differences.

These results suggest that training systems with Lombard speech is beneficial in terms of both estimated speech quality and estimated speech intelligibility. This is in line with and extends our preliminary study \citep{michelsanti2018effects}, where only a subset of the whole database was used to evaluate the models.

\begin{table}
\centering
\resizebox{0.45\textwidth}{!}{
\begin{tabular}{l | c c c c c c}
\toprule
PESQ &VO-L&VO-NL&AO-L&AO-NL&AV-L&AV-NL\\
\midrule
$-20$ - $5$ dB&1.163&1.113&1.353&1.283&1.446&1.331\\
\midrule
ESTOI &VO-L&VO-NL&AO-L&AO-NL&AV-L&AV-NL\\
\midrule
$-20$ - $5$ dB&0.372&0.335&0.448&0.423&0.528&0.488\\
\bottomrule
 \end{tabular}}
\caption{Average scores for the systems trained on a narrow SNR range.}
\label{tab:res_nar}
\end{table}

\subsubsection{Effects of Inter-Speaker Variability}

Previous work found a large inter-speaker variability for Lombard speech, especially between male and female speakers \citep{junqua1993lombard}. Here, we investigate whether this variability affects the performance of SE systems.

\begin{figure}
	\centering
		\includegraphics[scale=.5]{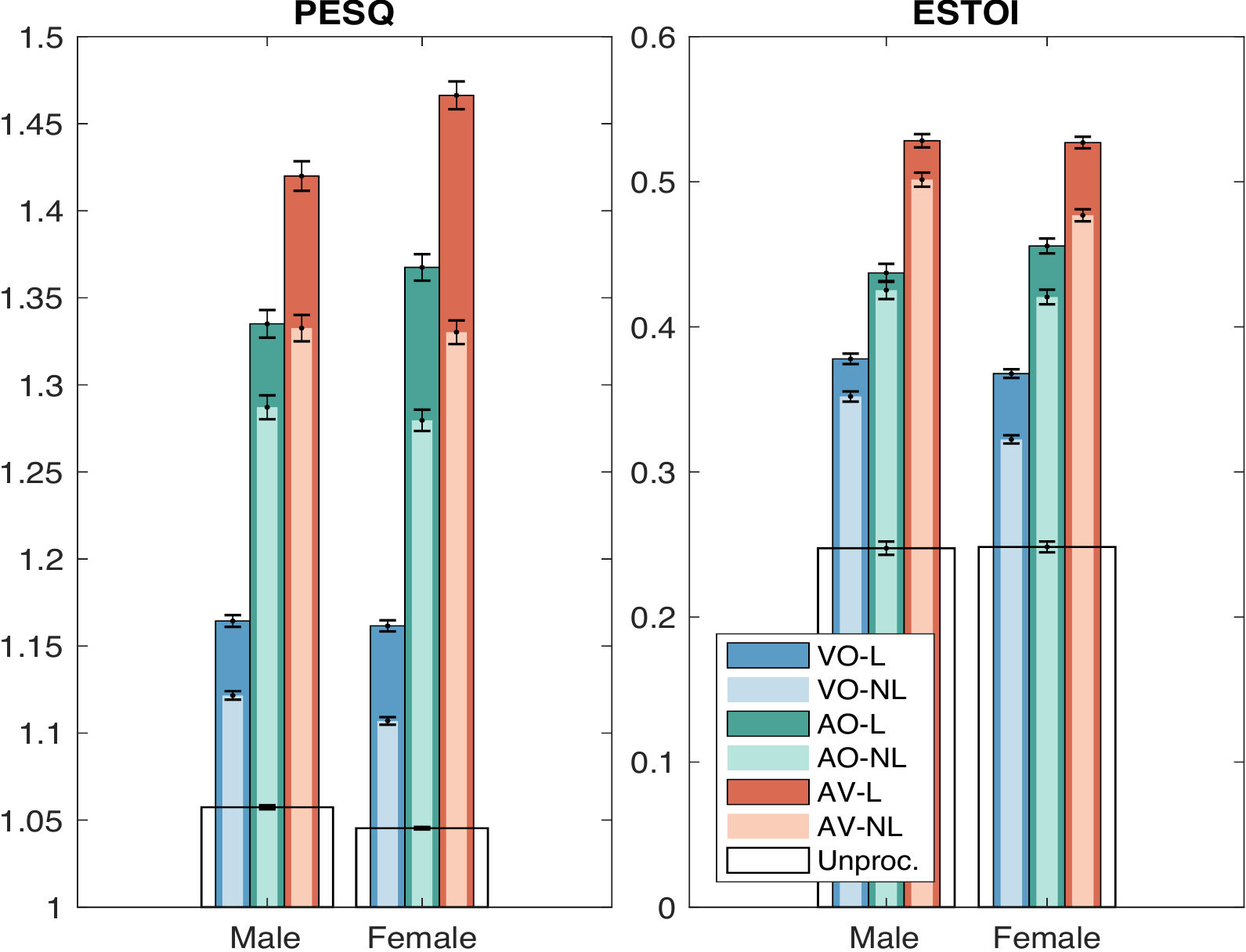}
	\caption{Cross-validation results for male and female speakers in terms of PESQ and ESTOI.}
	\label{fig:res_seen_gender}
\end{figure}

Figure \ref{fig:res_seen_gender} shows the average PESQ and ESTOI scores by gender. Since the scores are computed on different speech material, it may be hard to make a direct comparison between males and females by looking at the absolute performance. Instead, we focus on the gap between L and NL systems averaged across SNRs for same gender. At a first glance, the trends of the different conditions are as expected: L systems are better than the respective NL ones, and AV systems outperform AO systems trained with speech of the same speaking style, in terms of both estimated speech quality and estimated speech intelligibility. We also notice that the scores of VO systems are worse than the AO ones, also for ESTOI. This is because we average across all the SNRs and VO is better than AO only at very low SNRs, but considerably worse for $\text{SNR}\geq-5$ dB (Figure \ref{fig:res_seen}). 

The difference between L and NL systems is larger for females than it is for males. This can be observed for all the modalities and it is more noticeable for AV systems, most likely because they account for both audio and visual differences. In order to better understand this behaviour, we provide a more in-depth analysis, investigating the impact that some acoustic and geometric articulatory features have on estimated speech quality and estimated speech intelligibility.

\begin{figure}
	\centering
		\includegraphics[scale=.7]{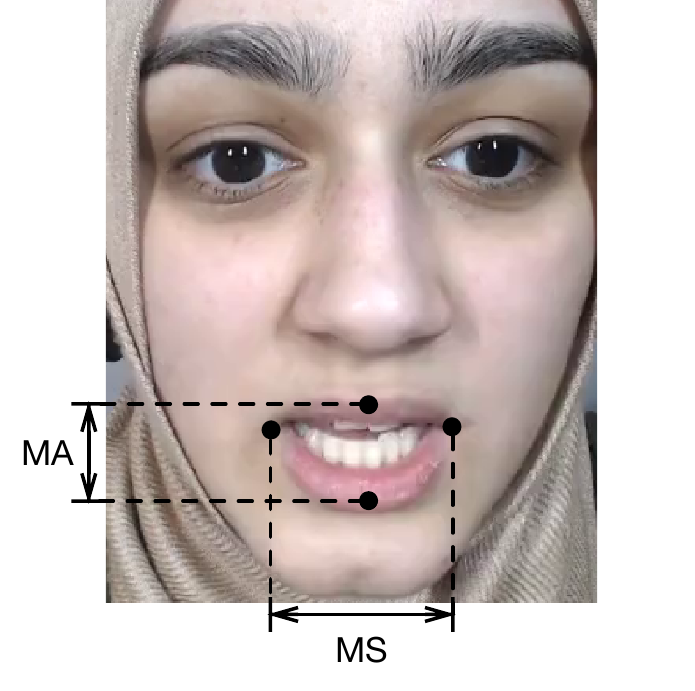}
	\caption{Mouth aperture (MA) and mouth spreading (MS) from $4$ facial landmarks.}
	\label{fig:visual_gestures}
\end{figure}

We consider three different features that have already been used to study Lombard speech in previous work \citep{garnier2006acoustic, garnier2012effect, tang2015examining, alghamdi2017visual}: F0, mouth aperture (MA) and mouth spreading (MS). The average F0 for each speaker was estimated with Praat \citep{boersma2001praat}, using the default settings for pitch estimation. The average MA and MS per speaker were computed from $4$ facial landmarks (Figure \ref{fig:visual_gestures}) obtained with the pose estimation algorithm \citep{kazemi2014one}, train\-ed on the iBUG 300-W database \citep{sagonas2016300}, implemented in the dlib toolkit \citep{dlib09}. 

\begin{figure*}
	\centering
		\includegraphics[scale=.45]{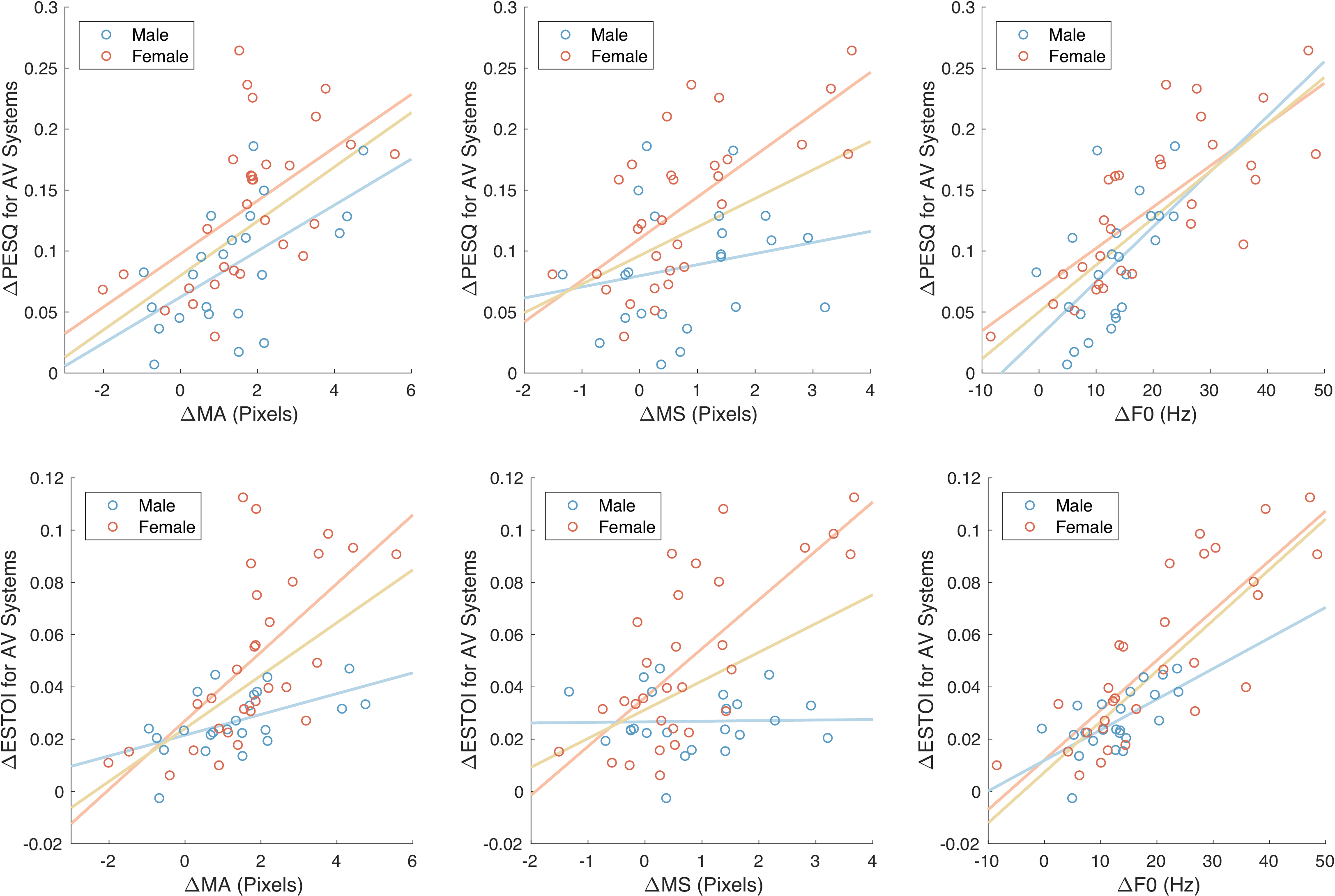}
	\caption{Scatter plots showing the relationship between the audio/visual features and PESQ/ESTOI. For each circle, which refers to a particular speaker, the $y$-coordinate indicates the average performance increment of AV-L with respect to AV-NL in terms of PESQ or ESTOI, while the $x$-coordinate indicates the average increment of audio (fundamental frequency) or visual (mouth aperture and mouth spreading) features in Lombard condition with respect to the respective feature in non-Lombard condition. The lines show the least-squares lines for male speakers (blue), female speakers (red), and all the speakers (yellow). MA, MS, and F0 indicate mouth aperture, mouth spreading, and fundamental frequency, respectively.}
	\label{fig:gestures}
\end{figure*}

Let $\Delta$F0, $\Delta$MA, and $\Delta$MS denote the average difference in audio and visual features, respectively, between Lombard and non-Lombard speech. Similarly, let $\Delta$PESQ and $\Delta$ESTOI denote the increment in PESQ and ESTOI, respectively, of AV-L with respect to AV-NL. Figure \ref{fig:gestures} illustrates the relationship between $\Delta$F0, $\Delta$MA, and $\Delta$MS and $\Delta$PESQ, and $\Delta$ESTOI. We notice that on average for each speaker $\Delta\text{PESQ}$ and $\Delta\text{ESTOI}$ are both positive, with only one exception represented by a male speaker, whose $\Delta\text{ESTOI}$ is slightly less than $0$. This indicates that no matter how different the speaking style of a person is in presence of noise, there is a benefit in training a system with Lombard speech. Focusing on the range of the features' variations, most of the speakers have positive $\Delta$MA, $\Delta$MS, and $\Delta$F0. This is in accordance with previous research, which suggests that in Lombard condition there is a tendency to amplify lips' movements and rise the pitch \citep{garnier2010influence, garnier2012effect, junqua1993lombard}. $\Delta$MA and $\Delta$MS values lie between $-2$ and $6$ pixels, and between $-2$ and $4$ pixels, respectively, for both male and female speakers. Regarding the $\Delta$F0 range, it is wider for females, up to $50$ Hz, against the $25$ Hz reached by males.

Among the three features considered, $\Delta$F0 is the one that seems to be related the most with $\Delta$PESQ and $\Delta$ESTOI. This can be seen by comparing the distributions of the circles with the least-squares lines in the plots of Figure \ref{fig:gestures} or by analysing the correlation between PESQ/ESTOI increments and audio/visual feature increments, using Pearson's and Spearman's correlation coefficients.

Given $n$ pairs of $(x_i, y_i)$ observations, with $i \in \{1, \dots, n\}$, from two variables $x$ and $y$, whose sample means are denoted as $\bar{x}$ and $\bar{y}$, respectively, we refer to the Pearson's correlation coefficient as $\rho_{P}(x,y)$.
%
%, is defined as \citep{sharma2005text}: 
%%
%\begin{equation}
%\rho_{P}(x,y) =\frac{\sum ^n _{i=1}(x_i - \bar{x})(y_i - \bar{y})}{\sqrt{\sum ^n _{i=1}(x_i - \bar{x})^2} \sqrt{\sum ^n _{i=1}(y_i - \bar{y})^2}},
%\label{eqn:pearsoncorr}
%\end{equation}
%%
%\noindent and it gives a measure of the linear relationship between $x$ and $y$ \citep{field2013discovering}.
We have that $-1 \leq \rho_{P}(x,y) \leq 1$, where $0$ denotes the absence of a linear relationship between the two variables, and $-1$ and $1$ a perfect positive linear relationship and a perfect negative linear relationship, respectively. To complement the Pearson's correlation coefficient, we also consider the Spearman's correlation coefficient, $\rho_{S}(x,y)$, defined as \citep{sharma2005text}:
\begin{equation}
\rho_{S}(x,y) = \rho_{P}(r_{x},r_{y}),
\label{eqn:spearmancorr}
\end{equation}
\noindent where $r_{x}$ and $r_{y}$ indicate rank variables. The advantage of using ranks is that $\rho_{S}$ allows to assess whether the relationship between $x$ and $y$ is monotonic (not limited to linear).

As shown in Table \ref{tab:corr_gestures}, for AV systems, $\Delta$F0 has a higher correlation with $\Delta$PESQ ($\rho_{P}=0.73$, $\rho_{S}=0.73$) and $\Delta$ESTOI ($\rho_{P}=0.81$, $\rho_{S}=0.77$) than $\Delta$MA and $\Delta$MS. We observe that for female speakers, the correlation between the features' increments and the performance measures' increments is usually higher, especially when considering $\Delta$MS, suggesting that some inter-gender difference should be present not only for $\Delta$F0 (whose range is way wider for females as previously stated), but also for visual features.

In Table \ref{tab:corr_gestures} we also report the correlation coefficients for the single modalities. The correlation of visual features' increments with $\Delta$PESQ or $\Delta$ESTOI is sometimes higher for AO systems than it is for VO systems. This might seem counter-intuitive, because AO systems do not use visual information. However, correlation does not imply causation \citep{field2013discovering}: since visual and acoustic features are correlated \citep{almajai2006analysis}, it is possible that other acoustic features, which are not considered in this study even though they might be correlated with $\Delta$MA and $\Delta$MS, play a role in the enhancement. Similar considerations can be done for $\Delta$F0, which has a correlation with $\Delta$ESTOI for VO systems ($\rho_{P}=0.77$, $\rho_{S}=0.77$) higher than the one for AO systems ($\rho_{P}=0.64$, $\rho_{S}=0.60$). By looking at the inter-gender differences, we find that, in general, the correlation coefficients computed for female speakers are higher than the ones computed for male speakers, especially when considering $\Delta$MS.

In general, a performance difference between genders exists when L systems are compared with NL ones, with a gap that is larger for females. This is unlikely to be caused by the small gender imbalance in the training set (23 males and 30 females). Instead, it is reasonable to assume that this result is due to the characteristics of the Lombard speech of female speakers, which shows a large increment of F0, the feature that correlates the most with the estimated speech quality and the estimated speech intelligibility increases, among the ones considered.

\begin{table}
\centering
\resizebox{0.45\textwidth}{!}{
\begin{tabular}{l | c c c | c c c }
\toprule
& &$\rho_{P}$ && & $\rho_{S}$ & \\
 & all & m & f & all & m & f\\
\midrule
$\Delta$PESQ (VO) - $\Delta$MA & $.29$ & $\:\:\:.32$ & $.24$ & $.35$ & $\:\:\:.30$ & $.29$\\
$\Delta$PESQ (AO) - $\Delta$MA & $.43$ & $\:\:\:.49$ & $.40$ & $.55$ & $\:\:\:.49$ & $.51$\\
$\Delta$PESQ (AV) - $\Delta$MA  & $.57$ & $\:\:\:.59$ & $.56$ & $.65$ & $\:\:\:.52$ & $.66$\\[3pt]
$\Delta$ESTOI (VO) - $\Delta$MA & $.46$ & $\:\:\:.19$ & $.57$ & $.52$ & $\:\:\:.16$ & $.69$\\
$\Delta$ESTOI (AO) - $\Delta$MA & $.43$ & $\:\:\:.47$ & $.46$ & $.52$ & $\:\:\:.52$ & $.50$\\
$\Delta$ESTOI (AV) - $\Delta$MA & $.57$ & $\:\:\:.53$ & $.65$ & $.67$ & $\:\:\:.47$ & $.72$\\
\midrule
$\Delta$PESQ (VO) - $\Delta$MS & $.19$ & $-.08$ & $.35$ & $.12$ & $-.03$ & $.31$\\
$\Delta$PESQ (AO) - $\Delta$MS & $.31$ & $\:\:\:.20$ & $.45$ & $.33$ & $\:\:\:.19$ & $.54$\\
$\Delta$PESQ (AV) - $\Delta$MS  & $.45$ & $\:\:\:.21$ & $.68$ & $.44$ & $\:\:\:.28$ & $.71$\\[3pt]
$\Delta$ESTOI (VO) - $\Delta$MS & $.45$ & $-.12$ & $.73$ & $.22$ & $-.21$ & $.62$\\
$\Delta$ESTOI (AO) - $\Delta$MS & $.30$ & $\:\:\:.05$ & $.47$ & $.22$ & $\:\:\:.07$ & $.48$\\
$\Delta$ESTOI (AV) - $\Delta$MS  & $.47$ & $\:\:\:.02$ & $.72$ & $.34$ & $-.02$ & $.66$\\
\midrule
$\Delta$PESQ (VO) - $\Delta$F0 & $.34$ & $\:\:\:.26$ & $.31$ & $.36$ & $\:\:\:.23$ & $.35$\\
$\Delta$PESQ (AO) - $\Delta$F0 & $.62$ & $\:\:\:.53$ & $.58$ & $.61$ & $\:\:\:.52$ & $.61$\\
$\Delta$PESQ (AV) - $\Delta$F0  & $.73$ & $\:\:\:.58$ & $.75$ & $.73$ & $\:\:\:.59$ & $.80$\\[3pt]
$\Delta$ESTOI (VO) - $\Delta$F0 & $.77$ & $\:\:\:.57$ & $.77$ & $.77$ & $\:\:\:.58$ & $.82$\\
$\Delta$ESTOI (AO) - $\Delta$F0  & $.64$ & $\:\:\:.55$ & $.60$ & $.60$ & $\:\:\:.56$ & $.61$\\
$\Delta$ESTOI (AV) - $\Delta$F0  & $.81$ & $\:\:\:.64$ & $.81$ & $.77$ & $\:\:\:.61$ & $.84$\\
\bottomrule
\end{tabular}}
\caption{Pearson's ($\rho_{P}$) and Spearman's ($\rho_{S}$) correlation coefficients between PESQ/ESTOI increments and audio/visual feature increments for male speakers (m), female speakers (f), and all the speakers. MA, MS, and F0 indicate mouth aperture, mouth spreading, and fundamental frequency, respectively.}
\label{tab:corr_gestures}
\end{table}

\subsection{Systems Trained on a Wide SNR Range}

The models presented in Section \ref{sec:narrowsnrmodels} have been trained to enhance signals when Lombard effect occurs, i.e. at SNRs between $-20$ and $5$ dB. However, from a practical perspective, SNR-independent systems, capable of enhancing both Lombard and non-Lombard speech, are preferred. There are several ways to achieve this goal. For example, it is possible to train a system (with Lombard speech) that works at low SNRs,  and another one (with non-Lombard speech) that works at high SNRs. This approach requires switching between the two systems, which can be problematic, because it involves an online estimation of the SNR. An alternative approach is to train general systems with Lombard speech at low SNRs and non-Lombard speech at high SNRs. We followed this alternative approach, building such systems and studying their strengths and limitations. We also compared them with systems trained only with non-Lombard speech for the whole SNR range, because this is what current state-of-the-art systems do.

\begin{figure*}
	\centering
		\includegraphics[scale=.52]{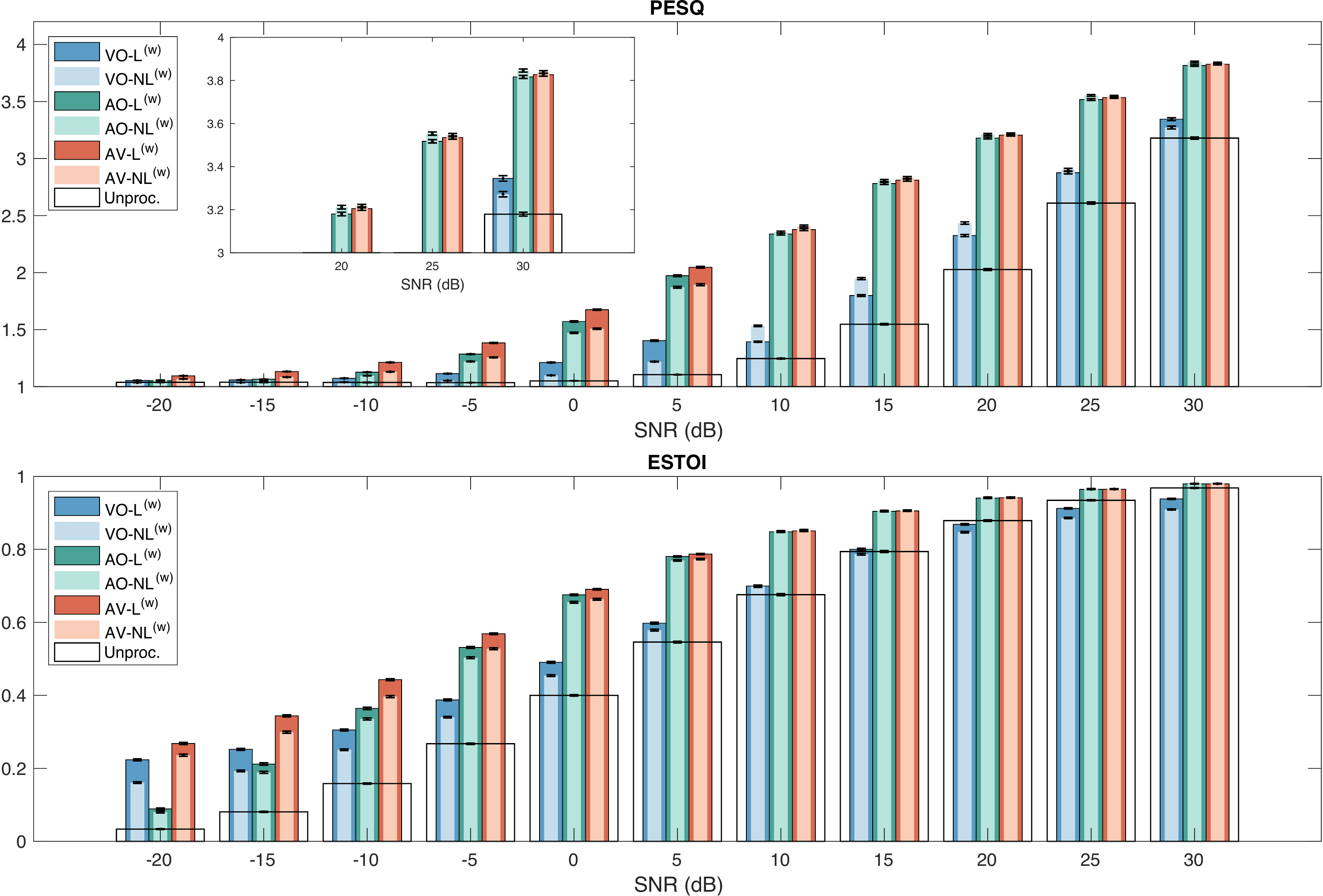}
	\caption{As Figure \ref{fig:res_seen}, but for the systems trained on a wide SNR range.}
	\label{fig:res_seen_g}
\end{figure*}

The test set was built by mixing additive SSN with Lombard speech at $6$ SNRs between $-20$ and $5$ dB, and with non-Lombard speech at $5$ SNRs between $10$ and $30$ dB. For VO-NL\textsuperscript{(w)}, AO-NL\textsuperscript{(w)}, and AV-NL\textsuperscript{(w)}, only non-Lombard speech was used during training, while for VO-L\textsuperscript{(w)}, AO-L\textsuperscript{(w)}, and AV-L\textsuperscript{(w)}, Lombard speech was used with $\text{SNR}\leq5$ dB and non-Lombard speech with $\text{SNR}\geq10$ dB, to match the speaking style of the test set (Table \ref{tab:models}). The results in terms of PESQ and ESTOI are shown in Figure \ref{fig:res_seen_g}. 

The relative performance of the systems at $\text{SNR} \leq 5$ dB is similar to the one observed for the systems trained on a narrow SNR range (Section \ref{sec:narrowsnrmodels}): L\textsuperscript{(w)} systems outperform the respective NL\textsuperscript{(w)} systems, AV performance is higher than AO and VO performance, and VO is considerably better than AO only in terms of ESTOI at very low SNRs.

When $\text{SNR}\geq10$ dB, NL\textsuperscript{(w)} systems perform better than L\textsuperscript{(w)} systems in terms of PESQ. The difference is on average (Table \ref{tab:res_wide}) larger for VO ($0.070$) than it is for AO ($0.028$) and AV ($0.018$). This can be explained by the fact that it is harder for VO-L\textsuperscript{(w)} to recognise when non-Lombard speech occurs using only the video of the speaker. However, these performance gaps are smaller than the ones between L\textsuperscript{(w)} and NL\textsuperscript{(w)} systems at $\text{SNR}\leq5$ dB ($0.073$ for VO, $0.051$ for AO, and $0.101$ for AV).

Regarding ESTOI at $\text{SNR}\geq10$ dB, the difference between AO and AV becomes negligible, with VO systems that perform considerably worse. This is because audio features are more informative than visual ones at high SNRs, making AO-SE systems already good to recover speech intelligibility. In addition, the average gaps between NL\textsuperscript{(w)} and L\textsuperscript{(w)} are quite small: $0.002$ for AO and AV, while for VO it is actually $-0.019$.

\begin{table}
\centering
\resizebox{0.48\textwidth}{!}{
\begin{tabular}{l | c c c c c c}
\toprule
PESQ &VO-L\textsuperscript{(w)}&VO-NL\textsuperscript{(w)}&AO-L\textsuperscript{(w)}&AO-NL\textsuperscript{(w)}&AV-L\textsuperscript{(w)}&AV-NL\textsuperscript{(w)}\\
\midrule
$-20$ - $\:\:5$ dB&1.153&1.080&1.346&1.295&1.424&1.323\\
$\:\:\:10$ - $30$ dB&2.348&2.418&3.127&3.155&3.151&3.169\\
\midrule
ESTOI &VO-L\textsuperscript{(w)}&VO-NL\textsuperscript{(w)}&AO-L\textsuperscript{(w)}&AO-NL\textsuperscript{(w)}&AV-L\textsuperscript{(w)}&AV-NL\textsuperscript{(w)}\\
\midrule
$-20$ - $\:\:5$ dB&0.376&0.330&0.442&0.422&0.517&0.483\\
$\:\:\:10$ - $30$ dB&0.844&0.825&0.927&0.929&0.928&0.930\\
\bottomrule
 \end{tabular}}
\caption{Average scores for the systems trained on a wide SNR range.}
\label{tab:res_wide}
\end{table}

In general, at $\text{SNR}\leq5$ dB, the systems that use both Lombard and non-Lombard speech for training perform better than the ones that only use non-Lombard speech. At higher SNRs, their PESQ and ESTOI scores are slightly worse than the ones of the systems trained only with non-Lombard speech. However, this performance gap is small, and seems to be larger for the estimated speech quality than for the estimated speech intelligibility. The way we combined non-Lombard and Lombard speech for training seems to be the best solution for an SNR-independent system, although a small performance loss may occur at high SNRs.

\section{Listening Tests}

Although it has been shown that visual cues have an impact on speech perception \citep{sumby1954visual, mcgurk1976hearing}, the currently available objective measures used to estimate speech quality and speech intelligibility, e.g. PESQ and ESTOI, only take into account the audio signals. Even when listening tests are performed to evaluate the performance of a SE system, visual stimuli are usually ignored and not presented to the participants \citep{hussain2017towards}, despite the fact that visual inputs are typically available during practical deployment of SE systems.

For these reasons, and in an attempt to evaluate the proposed AV enhancement systems in a setting as realistic as possible, we performed two listening tests, one to assess the speech quality and the other to assess the speech intelligibility, where the processed audio signals from the Lombard GRID corpus were accompanied by their corresponding visual stimuli. Both tests were conducted in a silent room, where a MacBookPro11,4 equipped with an external monitor, a sound card (Focusrite Scarlett 2i2) and a set of closed headphones (Beyerdynamic DT770) was used for audio and video playback. The multimedia player (VLC media player 3.0.4) was controlled by the subjects with a graphical user interface (GUI) modified from MUSHRAM \citep{vincent2005mushram}. The processed signals used in this test were from the systems trained on the narrow SNR range previously described (Section \ref{sec:narrowsnrmodels}). All the audio stimuli were normalised according to the two-pass EBU R128 loudness normalisation procedure \citep{ebu2014recommendation}, as implemented in ffmpeg-normalize\footnote{\url{https://github.com/slhck/ffmpeg-normalize}}, to guarantee that signals of different conditions were perceived as having the same volume. The subjects were allowed to adjust the general loudness to a comfortable level during the training session of each test.

\subsection{Speech Quality Test}

The quality test was carried out by $13$ experienced listeners, who volunteered to be part of the study. The participants were between 26 and 44 years old, and had self-reported normal hearing and normal (or corrected to normal) vision. On average, each participant spent approximately $30$ minutes to complete the test.

\subsubsection{Procedure}

The test used the MUlti Stimulus test with Hidden Reference and Anchor (MUSHRA) \citep{itu2003recommendation} paradigm to assess the speech quality on a scale from $0$ to $100$, divided into $5$ equal intervals labelled as \textit{bad}, \textit{poor}, \textit{fair}, \textit{good}, and \textit{excellent}. No definition of \textit{speech quality} was provided to the participants. Each subject was presented with $2$ sequences of $8$ trials each, $4$ to evaluate the systems at $-5$ dB SNR, and $4$ to evaluate the systems at $5$ dB SNR. Lower SNRs were not considered to ensure that the perceptual quality assessment was not influenced too much by the decrease in intelligibility. One trial consisted of one reference (clean speech signal) and seven other signals to be rated with respect to the reference: $1$ hidden reference, $4$ systems under test (AO-L, AO-NL, AV-L, AV-NL), $1$ unprocessed signal, and $1$ hidden anchor (unprocessed signal at $-10$ dB SNR). The participants were allowed to switch at will between any of the signals inside the same trial. The order of presentation of both the trials and the conditions was randomised, and signals from $4$ different randomly chosen speakers were used for each sequence of trials.

Before the actual test, the participants were trained in a special separate session, with the purpose of exposing them to the nature of the impairments and making them familiar with the equipment and the grading system.

\subsubsection{Results and Discussion}
\label{subsec:mushra_res}

\begin{figure}
	\centering
		\includegraphics[scale=.5]{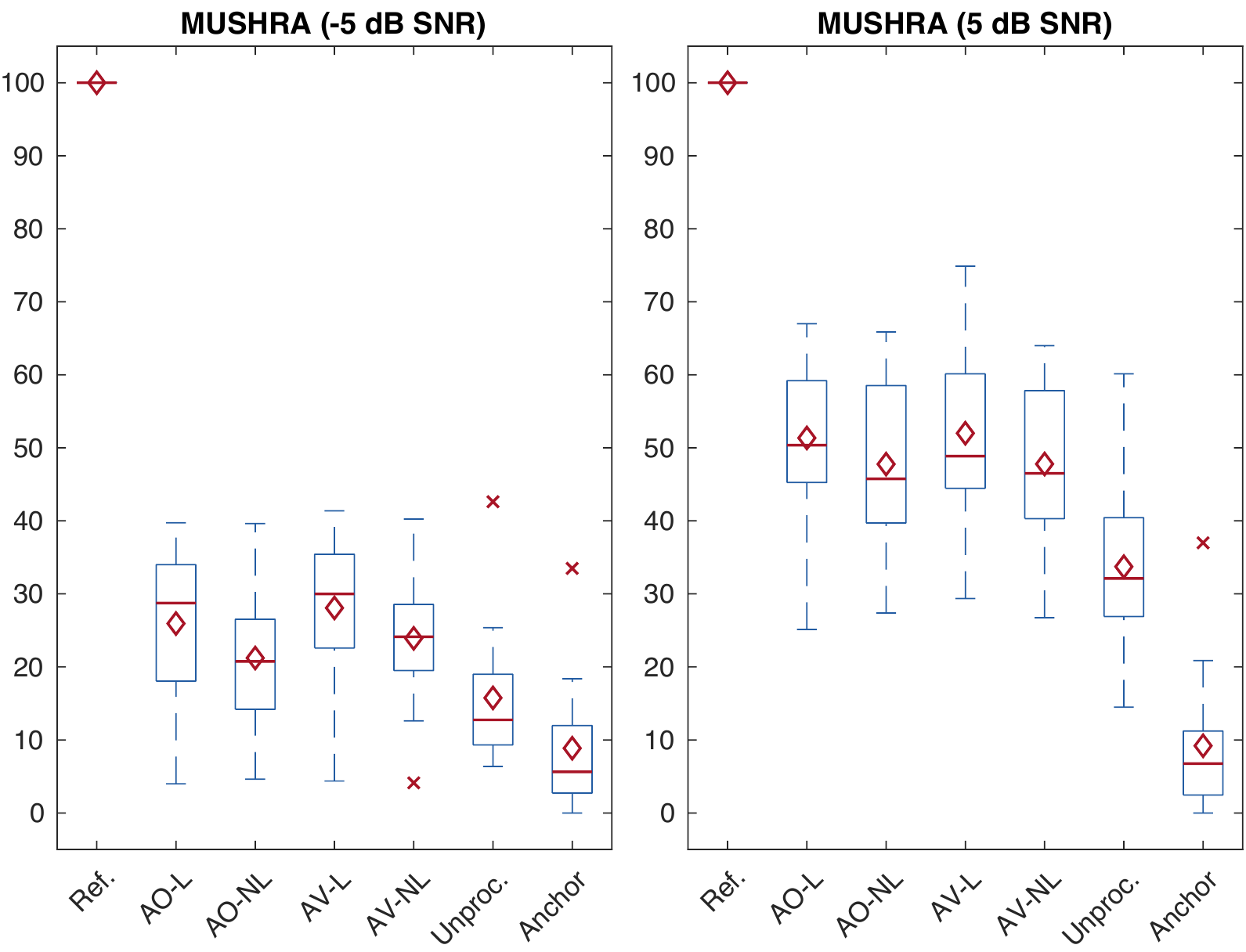}
	\caption{Box plots showing the results of the MUSHRA experiments for the signals at $-5$ dB SNR (left) and at $5$ dB SNR (right). The red horizontal lines and the diamond markers indicate the median and the mean values, respectively. Outliers (identified according to the 1.5 interquartile range rule) are displayed as red crosses. \textit{Ref.} indicates the reference signals.}
	\label{fig:mushra}
\end{figure}

The average scores assigned by the subjects for each condition are shown in Figure \ref{fig:mushra} in the form of box plots. 
 
Non-parametric approaches are used to analyse the data \citep{mendoncca2018statistical, winter2018colouration}, since the assumption of normal distribution of the data is invalid, given the number of participants and their different interpretation of the MUSHRA scale. Specifically, the paired two-sided Wilcoxon signed-rank test \citep{wilcoxon1945individual} is a\-dopted to determine whether there exists a median difference between the MUSHRA scores obtained for two different conditions. Differences in median are considered significant for $p < \alpha / m = 0.0083$ ($\alpha=0.05$, $m=6$), where the significance level is corrected with the Bonferroni method to compensate for multiple hypotheses tests \citep{field2013discovering}. The use of $p$-values as the only analysis strategy has been heavily criticized \citep{hentschke2011computation} because statistical significance can be obtained with a big sample size \citep{sullivan2012using, moore2012introduction} even if the magnitude of the effect is negligible \citep{hentschke2011computation}. For this reason, we complement $p$-values with a non-parametric measure of the effect size, the Cliff's delta \citep{cliff1993dominance}:
\begin{equation}
d_{C} = \frac{\sum_{i=1}^{m} \sum_{j=1}^{n} [x_i > y_j] - \sum_{i=1}^{m} \sum_{j=1}^{n}[x_i < y_j]}{mn},
\label{eqn:cliffdelta}
\end{equation}
\noindent where $x_i$ and $y_j$ are the observations of the samples of sizes $m$ and $n$ to be compared and $[P]$ indicates the Iverson bracket, which is $1$ if $P$ is true and $0$ otherwise. %We implemented $d_{C}$ using the area under the receiver operating characteristic curve (AUROC) in the MATLAB toolbox developed by \citet{hentschke2011computation} as $d_{C} = 2\, \text{AUROC}-1$ \citep{kraemer2006size}. 
As reported in Table \ref{tab:eff_size}, we consider the effect size to be small if $0.11 \leq | d_{C} | < 0.28$, medium if $ 0.28 \leq | d_{C} | < 0.43 $, and large if $|d_{C}| \geq 0.43$, according to the indication by \citet{vargha2000critique}. The $p$-values and the effect sizes for the comparisons considered in this study are shown in Table \ref{tab:mushra_pval}.

\begin{table}
\centering
%\resizebox{0.45\textwidth}{!}{
\begin{tabular}{c c c }
\toprule
Small Effect Size & Medium Effect Size    & Large Effect Size   \\
\midrule
$0.11 \leq | d_{C} | < 0.28$       &      $ 0.28 \leq | d_{C} | < 0.43 $		&    $0.43 \leq |d_{C}| \leq 1$ \\ 
\bottomrule
\end{tabular}%}
\caption{Interpretation of the effect size (Cliff's delta, $d_{C}$). Adapted from \citep{vargha2000critique}.}
\label{tab:eff_size}
\end{table}

\begin{table}
\centering
%\resizebox{0.45\textwidth}{!}{
\begin{tabular}{l | c c | c c }
\toprule
                           \multicolumn{1}{c}{}   & \multicolumn{2}{c}{$-5$ dB SNR} & \multicolumn{2}{c}{$5$ dB SNR} \\
\midrule
{Comparison} & $p$   & $d_{C}$   & $p$   & $d_{C}$   \\
\midrule
{AO-L - AO-NL}       &      $<.0083$		&    $\:\:\:.30$      	&     $\quad.0134$	&   $\:\:\:.22$        \\ 
{AV-L - AV-NL}         &      $<.0083$        	&    $\:\:\:.32$      	&     $<.0083$		&   $\:\:\:.23$          \\ 
{AO-L - AV-L}           &      $\quad.0498$     &    $- .14$      		&     $\quad.7476$	&   $\:\:\:.02$        \\ 
{AO-NL - AV-NL}      &      $<.0083$         	&    $-.21$      		&     $\quad.8262$	&   $-.02$     \\ 
{AO-L - Unproc.}      &      $\quad.0479$	&    $\:\:\:.57$   		&     $<.0083$		&   $\:\:\:.74$      \\ 
{AV-L - Unproc.}       &      $\quad.0134$	&    $\:\:\:.59$   		&     $<.0083$		&   $\:\:\:.79$     \\ 
\bottomrule
\end{tabular}%}
\caption{$p$-values ($p$) and effect sizes (Cliff's delta, $d_{C}$)  for the MUSHRA experiments. The significant level (0.0083) for the $p$-values is corrected with the Bonferroni method.}
\label{tab:mushra_pval}
\end{table}

At $\text{SNR} = -5$ dB, a  significant ($p < 0.0083$) medium ($0.28 < |d_{C}| < 0.43$) difference exists between Lombard and non-Lombard systems for both the audio-only and the audio-visual cases. The increment in quality when using vision with respect to audio-only systems is perceived by the subjects ($| d_{C} | > 0.11$), but it has only a relatively small effect ($| d_{C} | < 0.28$). This was expected, since visual cues affect more the intelligibility at low SNRs than quality, as also shown by objective measures (Figure \ref{fig:res_seen}). More specifically, for non-Lombard systems, this difference is significant and greater than the one found for Lombard systems, meaning that vision contributes more when the enhancement of Lombard speech is performed with systems that were not trained with it. We can notice that there is a large ($|d_{C}| > 0.43$) difference between the unprocessed signals and the version enhanced with Lombard systems. However, this difference is not significant, probably due to the heterogeneous interpretation of the MUSHRA scale by the subjects and their preference of the different natures of the impairment (presence of noise or artefacts caused by the enhancement).

At an SNR of 5 dB a small difference between Lombard and non-Lombard systems is observed, despite being not significant in the audio-only case ($p=0.0134$). At this noise level, audio-visual systems appear to be indistinguishable ($ |d_{C}| < 0.11$) from the respective audio-only systems. This confirms the intuition that vision does not help in improving the speech quality at high SNRs. Finally, the difference between the unprocessed signals and the respective enhanced versions using Lombard systems is both large ($|d_{C}| > 0.43$) and significant ($p < 0.0083$), which makes it clear that both AO-L and AV-L improve the speech quality.

\subsection{Speech Intelligibility Test}

The intelligibility test was carried out by $11$ listeners, who volunteered to be part of the study. The participants were between 24 and 65 years old, and had self-reported normal hearing and normal (or corrected to normal) vision. On average, each participant spent approximately $45$ minutes to complete the test.

\subsubsection{Procedure}

Each subject was presented with $2$ sequences of $80$ audio-visual stimuli from the Lombard GRID corpus: $8$ speakers $\times$ $4$ SNRs ($-20$, $-15$, $-10$, and $-5$ dB) $\times$ $5$ processing conditions (unprocessed, AO-L, AO-NL, AV-L, AV-NL). The participants were asked to listen to each stimulus only once and, based on what they heard, they had to select the colour and the digit from a list of options and to write the letter (Table \ref{tab:syntax}). The order of presentation of the stimuli was randomised.

Before the actual test, the participants were trained in a special separate session consisting of a sequence of $40$ audio-visual stimuli.

\subsubsection{Results and Discussion}

\begin{figure*}
	\centering
		\includegraphics[scale=.65]{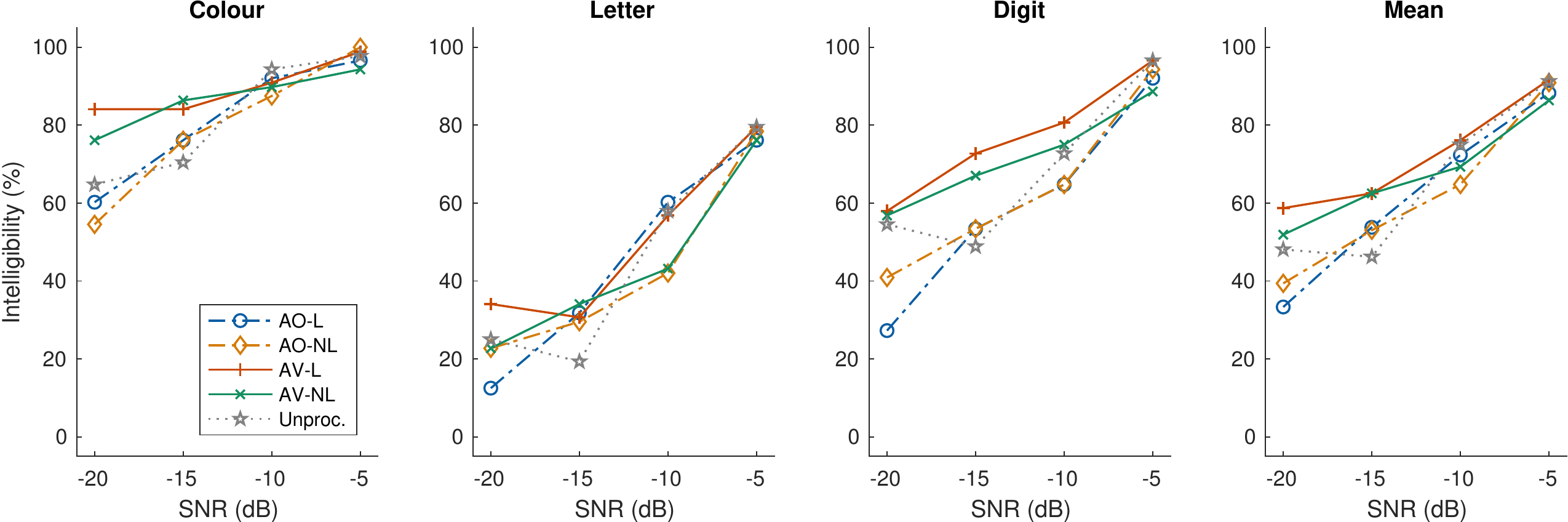}
	\caption{Percentage of correctly identified words obtained in the listening tests for the colour, the letter, and the digit fields, averaged across $11$ subjects. The mean intelligibility scores for all the fields are also reported.}
	\label{fig:intell_res}
\end{figure*}

The mean percentage of correctly identified keywords as a function of the SNR is shown in Figure \ref{fig:intell_res}. We can see that among the three fields, the colour is the easiest word to be identified by the participants. In general, the following trends can be observed. At low SNRs the intelligibility of the signals enhanced with AV systems is higher than the intelligibility obtained with AO systems. This difference substantially diminishes when the SNR increases. There is no big performance difference between L and NL systems, but in general AV-L tends to have higher percentage scores than the other systems. AV-L is also the only system that does not decrease the mean intelligibility scores for all the fields if compared to the unprocessed signals.

\begin{table}
\centering
%\resizebox{0.45\textwidth}{!}{
\begin{tabular}{l | c c c c }
\toprule
$p$ & \multicolumn{4}{c}{SNR} \\
\midrule
Comparison& -20 dB& -15 dB& -10 dB& -5 dB\\
\midrule
AO-L - AO-NL&$.3066$&$.4688$&$.0430$&$.2539$\\
AV-L - AV-NL&$.0625$&$.8633$&$.0742$&$.1055$\\
AO-L - AV-L&$.0010$&$.0117$&$.5625$&$.2344$\\
AO-NL - AV-NL&$.0527$&$.0430$&$.3359$&$.2070$\\
AO-L - Unproc.&$.0332$&$.0547$&$.9004$&$.1250$\\
AV-L - Unproc.&$.1270$&$.0078$&$.8828$&$.8828$\\
\midrule
$d_{C}$ & \multicolumn{4}{c}{SNR} \\
\midrule
Comparison& -20 dB& -15 dB& -10 dB& -5 dB\\
\midrule
AO-L - AO-NL&$-.08$&$\:\:\:.06$&$\:\:\:.31$&$-.31$\\
AV-L - AV-NL&$\:\:\:.32$&$\:\:\:.01$&$\:\:\:.39$&$\:\:\:.28$\\
AO-L - AV-L&$-.91$&$-.35$&$-.17$&$-.34$\\
AO-NL - AV-NL&$-.32$&$-.37$&$-.31$&$\:\:\:.21$\\
AO-L - Unproc.&$-.31$&$\:\:\:.17$&$-.09$&$-.26$\\
AV-L - Unproc.&$\:\:\:.18$&$\:\:\:.46$&$0$&$\:\:\:.08$\\
\bottomrule
\end{tabular}%}
\caption{$p$-values ($p$) and effect sizes (Cliff's delta, $d_{C}$)  for the mean intelligibility scores for all the keywords obtained in the listening tests.}
\label{tab:intell_tab}
\end{table}

Table \ref{tab:intell_tab} shows Cliff's deltas and $p$-values, computed with the paired two-sided Wilcoxon signed-rank test, as in the MUSHRA experiments. 

The effect sizes support the observations made from Figure \ref{fig:intell_res}. Medium and large differences ($| d_{C} | > 0.28$) exist between AO and AV systems, especially at low SNRs. While AO-L and AO-NL are indistinguishable ($|d_{C}| < 0.11$) for SNR $<-10$ dB, there is a medium ($ 0.28 \leq | d_{C} | < 0.43 $) difference between AV-L and AV-NL, except for $-15$ dB SNR ($d_{C}=0.01$). Moreover, the intelligibility increase of AV-L over the unprocessed signals is perceived by the subjects at SNR $\leq-15$ dB ($| d_{C} | > 0.11$).

Regarding the $p$-values, if we focus on each SNR separately, the difference between two approaches can be considered significant for $p < 0.0083$ (cf. Section \ref{subsec:mushra_res}). This condition is met only when we compare AO-L with AV-L at $-20$ dB SNR and AV-L with the noisy speech at $-15$ dB SNR.

There are three main sources of variability that most like\-ly prevent the differences to be significant. First, the variation in lipreading ability among individuals is large and does not directly reflect the variation found in auditory speech perception skills \citep{summerfield1992lipreading}. Secondly, individuals have very different fusion responses to discrepancy in the auditory and visual syllables \citep{mallick2015variability}, which in our case might occur due to the artefacts produced in the enhancement process. Finally, the participants were not exposed to the same utterances processed with the different approaches like in MUSHRA. Since the vocabulary set of the Lombard GRID corpus is small and some words are easier to understand because they contain unambiguous visemes, the intelligibility scores are affected not only by the various processing conditions, but also by the different sentences used.

\section{Conclusion}

In this paper, we presented an extensive analysis of the impact of Lombard effect on audio, visual and audio-visual speech enhancement systems based on deep learning. We conducted several experiments using a database consisting of $54$ speakers and showed the general benefit of training a system with Lombard speech.

In more detail, we first trained systems with Lombard or non-Lombard speech and evaluated them on Lombard speech adopting a cross-validation setup. The results showed that systems trained with Lombard speech outperform the respective systems trained with non-Lombard speech in terms of both estimated speech quality and estimated speech intelligibility. We also observed a performance difference across speakers, with an evident gap between genders: the performance difference between the systems trained with Lombard speech and the ones trained with non-Lombard speech is larger for females than it is for males. The analysis that we performed suggests that this difference might be primarily due to the large increment in the fundamental frequency that female speakers exhibit from non-Lombard to Lombard conditions.

With the objective of building more general systems able to deal with a wider SNR range, we then trained systems using Lombard and non-Lombard speech and compared them with systems trained only on non-Lombard speech. As in the narrow SNR case, systems that include Lombard speech perform considerably better than the others at low SNRs. At high SNRs, the estimated speech quality and the estimated speech intelligibility obtained with systems trained only with non-Lombard speech are higher, even though the performance gap is very small for the audio and the audio-visual cases. Combining non-Lombard and Lombard speech for training in the way we did guarantees a good compromise for the enhancement performance across all the SNRs.

We also performed subjective listening tests with audio-visual stimuli, in order to evaluate the systems in a situation closer to the real-world scenario, where the listener can see the face of the talker. For the speech quality test, we found significant differences between Lombard and non-Lombard systems at all the used SNRs for the audio-visual case and only at $-5$ dB SNR for the audio-only case. Regarding the speech intelligibility test, we observed that on average the scores obtained with the audio-visual system trained with Lombard speech are higher than the other processing conditions. However, we were unable to find significant differences in most of the cases, suggesting that in future works more effort should be put into designing new paradigms for speech intelligibility tests to control the several sources of variability caused by the combination of auditory and visual stimuli.

%{\color{blue}Assumptions and limitations of the experiments:
%1. We artificially added noise to NL and L speech.
%2. We assumed that the Lombard effect is the same across all the SNRs (between -20 and 5 dB).
%3. The dataset is in controlled conditions with a small vocabulary set.}

\section{Acknowledgements}

This work was supported, in part, by the Oticon Foundation.

\bibliography{cas-dc-test}

\makeatletter

\def\pct{\expandafter\@gobble\string\%}

\immediate\write\@auxout{\pct\space This is a test line.\pct }

\end{document}